\documentclass[%
 reprint,
amsmath,amssymb,
 aps,
]{revtex4-1}

\usepackage{graphicx}
\usepackage{dcolumn}
\usepackage{bm}
\usepackage{epsfig,graphics,amsmath,amssymb}
\usepackage{epstopdf}
\usepackage{upgreek}
\usepackage{ragged2e}
\usepackage{titlesec}
\usepackage{caption}
\usepackage{float}
\usepackage{setspace}
\titleformat*{\section}{\normalsize\bfseries\rmfamily}

\begin{document}

\preprint{APS/123-QED}

\title{Crystal Symmetry and  Polarization of High-order Harmonics in ZnO}

\author{Shicheng Jiang$^{1}$, Shima Gholam-Mirzaei$^{2}$, Erin Crites$^{2}$, John E. Beetar$^{2}$, Mamta Singh$^{2,3}$, Ruifeng Lu$^{1,\dagger}$, Michael Chini$^{2,3,*}$ and C.~D.~Lin$^{4}$}

\affiliation{$^{1}$Department of Applied Physics, Nanjing University of Science and Technology, Nanjing 210094, P. R. China}
\affiliation{$^{2}$Department of Physics, University of Central Florida, Orlando, Florida 32816, USA}
\affiliation{$^{3}$CREOL, the College of Optics and Photonics, University of Central Florida, Orlando, Florida 32816, USA,}
\affiliation{$^{4}$J. R. Macdonald Laboratory, Department of Physics, Kansas State University, Manhattan, Kansas 66506, USA}

\email[Email:]{$^{*}$Michael.Chini@ucf.edu}
\email[Email:]{$^{\dagger}$ rflu@njust.edu.cn}

\date{\today}

\begin{abstract}
    We carried out a joint theoretical and experimental study of the polarization of high-order harmonics generated from ZnO by intense infrared laser pulses.   Experimentally we found that the dependence of parallel and perpendicular polarizations on the crystal orientation for all odd harmonics are nearly identical, but they are quite different from even harmonics which also show little order dependence. A one-dimensional two-band model, combined with a linear coupled excitation model, is shown to be able to explain the observed polarization behavior, including low-order harmonics. We further note that the same odd/even order contrast have been reported in a number of other crystals, despite that the harmonics were perceived to be generated via entirely different mechanisms. We demonstrated that this universality is governed by crystal symmetry, not by specific mechanisms.  Thus, polarization measurements  of harmonics offers a powerful pure optical method for determining the crystal axes as well as   monitoring their ultrafast changes when crystals are undergoing deformation. In addition, the ellipticity of harmonic has been studied. It shows that ellipticity of high-order harmonics from solids can be tuned precisely by changing the bond structure of the sample.

\end{abstract}

\maketitle

\section{Introduction}
High-order harmonics generated by intense laser pulses in gases have been investigated over the last three decades as a means for providing new extreme ultraviolet and soft X-ray light sources. They are also responsible for the emergence of attosecond science\cite{attopulse-1,attopulse-2}. When harmonics are generated in the molecular frame (i.e., fixed-in-space), the anisotropy and the inherent spatial symmetry of the electronic structure of the molecule are expected to result in polarized harmonic radiation. Since gas-phase molecules can only be partially oriented, the polarization states of harmonics from molecules have been studied only rarely \cite{levesque,jila,Bordeau, ATLpol}. High-order harmonic generation (HHG) from solids has a much shorter history. The first experiment  demonstrated HHG was from ZnO\cite{ZnO-2011} using mid-infrared laser pulses.   Since then, harmonics generated from various crystals have been reported \cite{chini_apl,Vampa_nature,Vampa_prl,ohio_nc,shambhu_mgo,shambhu_JPB2014,
shambhu_OL,T.T.Lu_nature2015,Garg_nature2016,shambhu_nc2017,
schubert_naturephoton2014,Hohenleutne_nature2015,Langer_naturephoton2017,
shambhu_mos2,graphene_science2017,Taucer_prb,berrycurvature-TTLu,GaSe-PRL,
Shambhu-MgO-2018,Shima,solidargon,A.A.Lanin_optica2017}, focusing largely on the dependence of harmonic yields on laser intensity and/or orientation of the crystal axis \cite{chini_apl,shambhu_mgo,schubert_naturephoton2014,Langer_naturephoton2017,shambhu_mos2,berrycurvature-TTLu,GaSe-PRL,Shambhu-MgO-2018}. Several experiments have also reported the polarization properties of the harmonics \cite{Langer_naturephoton2017,
shambhu_mos2,berrycurvature-TTLu,GaSe-PRL}.

HHG from gases can be qualitatively understood using the three-step model of ionization, acceleration, and recollision \cite{threestep1, threestep2,threestep3}. This model has been extended to the quantitative rescattering (QRS) theory \cite{cdlbook,ATLQRS}, which allows the extraction of photo-recombination transition dipole matrix elements from experimental spectra. The recombination dipole is related to the photoionization transition dipole, which reflects the molecular symmetry. Such symmetry would appear in the polarization of harmonics as well; however in experiments it is blurred by molecular rotation. For harmonics from solids, the crystal axes can be precisely aligned without rotational motion, and thus experimentally-obtained harmonic polarization states are expected to reflect the crystal symmetry. Indeed, experimental harmonic spectra are observed to reveal rotational symmetry \cite{shambhu_mgo,Shambhu-MgO-2018}. However, the role of other symmetries in solids has been less studied. Here, we investigate the role of crystal symmetry properties in determining the polarization states of HHG.

For solids, time-dependent density functional theory (TDDFT)\cite{TDDFT-original, Rubio, floss}, time-dependent Schr\"odinger equations (TDSE) \cite{{WMX_pra2015}} and Semiconductor Bloch Equations (SBEs) \cite{Golde_prb2008,ttluu-prb} are usually used to calculate harmonics generated from solids.  A three-step model similar to the case of gaseous media  was also developed for two-band solid system \cite{solid-sfa}. The SBEs method is often favored since it offers a simple way to account for dephasing of electron trajectories. We are aware   that   Floss et al. has extended the TDDFT to open quantum system taking dephasing effect into account, but such calculations are much more complicated \cite{TDDFT-open quantum system}.  Thus, in practical applications, one-dimensional (1D) SBEs are often employed to calculate HHG from solids. Depending on the band structure of the solid, the number of valence bands and conduction bands in the SBEs calculations can be varied. In general, multiband calculations are difficult to analyze, but this is not the case for ZnO which is dominated by two bands. Fig. 1 summarizes the structure of ZnO in real space. HHG from ZnO was first studied in 2011\cite{ZnO-2011} and again in 2016\cite{chini_apl}. In both experiments, the dependence of harmonic spectra on crystal orientation with respect to the driving laser polarization direction were reported. Both even and odd harmonics are clearly seen in the experiments, which is consistent with the lack of inversion symmetry along the ZnO c-axis. However, orientation-dependent features of even harmonics in ZnO could not be reproduced theoretically until recently, when Jiang et al. \cite{jiang_pra,jiang_prl} took the transition dipole phase into consideration.

Previous 1D calculations by Jiang et al. \cite{jiang_prl} only obtain harmonics polarized in the same direction as the driving laser polarization. However, most measurements have not discriminated between parallel and perpendicularly polarized harmonics. While it is straightforward to extend the SBEs model to 2D or 3D,  the numerical effort is large. Interestingly, a simple Linearly Coupled Excitation (LCE) model has been proposed previously by Koch and collaborators \cite{Langer_naturephoton2017}. The model has been applied to calculate   parallel and perpendicularly polarized harmonics in GaSe generated by THz pulses. The method involves solving the 1D SBEs for the induced current along the direction of each bond in the crystal. These bonds of course underlie the intrinsic symmetry of the crystal. The LCE model was implemented by Jiang et al\cite{jiang_prl} to obtain parallel and perpendicular harmonics\cite{comment} in ZnO. By including the two harmonic polarization components, improved agreement with the experimental data of Ref. 8 was observed. Still, a critical test of the model is to compare with polarization-sensitive experiments. By carrying out the LCE model calculation in conjunction with polarization-sensitive measurements, we report in this paper that the polarization states of harmonics can largely be explained by symmetry, using the LCE model.

\begin{figure}
    \includegraphics[width=3.5in,angle=0]{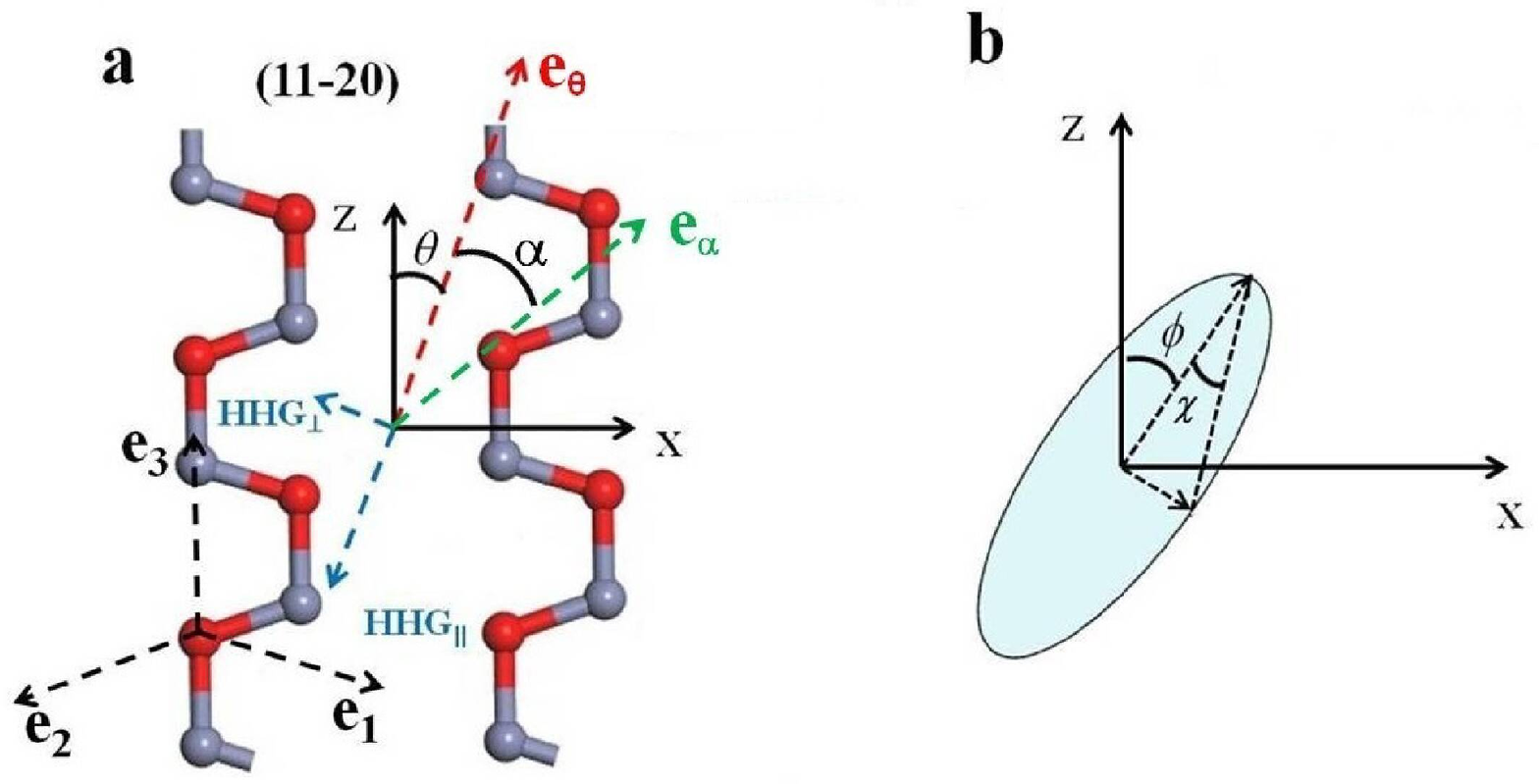}
   \caption* \justifying{Fig. 1. (a) The arrangement of atoms in real space on the (11-20) plane, red for oxygen and gray for zinc atoms. Three unit vectors ${\bf{e_1}}$, ${\bf{e_2}}$ and ${\bf{e_3}}$ are defined along the bond directions. The polarization of the driving laser makes an angle ${\theta}$ with respect to the crystal axis (along z). The polarization angle of the harmonic makes an angle $\alpha$ with respect to the polarization of the driving laser. (b) Definition of the polarization ellipse and the angle ${\chi}$ which is related to the ellipticity by $\epsilon$ = $tan({\chi}).$}\label{Fig_1} \justifying
\end{figure}

\begin{figure}
\centering
    \includegraphics[width=3.5in,angle=0]{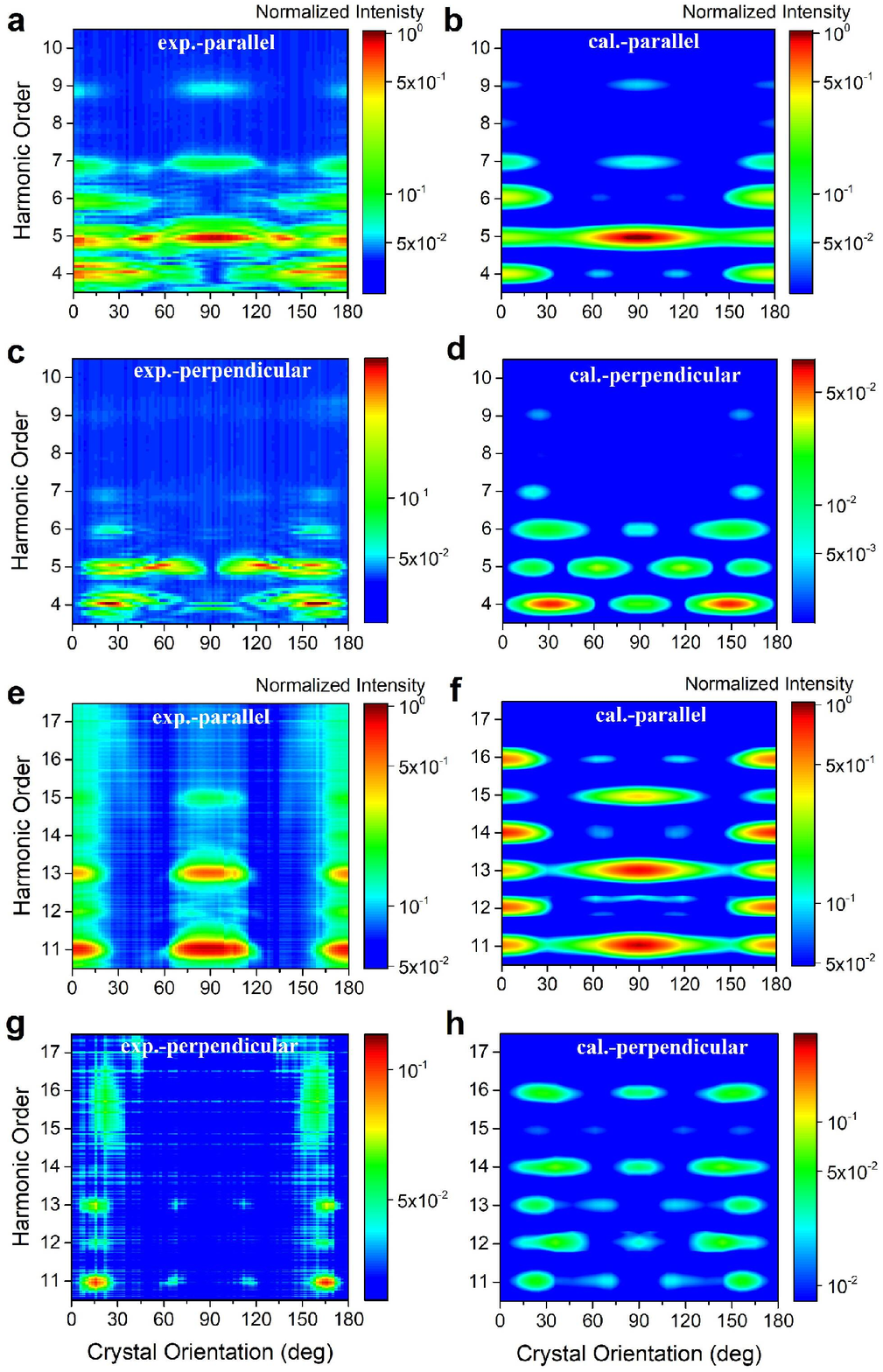}
    \caption* \justifying{Fig. 2. The left column shows the experimental results and the right column shows the simulations. The top two rows are for below-gap harmonics and the bottom two rows are for above-gap harmonics. The first row of each group is for parallel polarization component and the second row is for perpendicular polarization component. The simulation for each harmonic is normalized to experimental data separately for best visual agreement. A single scaling factor is used for both polarizations and all crystal orientations. Weaker features in above-gap harmonics in the experiment lie below the noise level.
}\label{Fig_2}\justifying
\end{figure}
\section{Polarization properties of HHG}
 In the experiments, harmonics were produced in ZnO using a commercial optical parametric amplifier (Light Conversion ORPHEUS). The idler pulses, with central wavelength of 3.6 ${\mu}$m, energy of 10 ${\mu}$J, and pulse duration of 90 fs were focused onto the exit plane of a 0.3-mm thick a-cut zinc oxide crystal. The crystal could be rotated about its surface normal, to vary the orientation of the c-axis with respect to the laser polarization direction. The generated harmonics passed through a polarizer and were detected using a UV-enhanced spectrometer (Ocean Optics HR2000+ES) capable of detecting harmonics from 4th to 17th order.
Two types of measurements were performed. In the first case, we fixed the angle between the laser polarization and the c-axis, and rotated the polarizer to analyze the polarization states of the generated harmonics. In the second set of experiments, we fixed the polarizer and varied the crystal orientation. In this case, the polarizer was set to pass harmonics generated with polarization parallel or perpendicular to the driving laser polarization, and the harmonic spectra were measured as a function of the crystal orientation. All these measurements were done using two different polarizers: a sheet polarizer (400-750 nm) was used to measure the polarization states of below-gap harmonics with photon energy below the ZnO band gap (3.3 eV), while a broadband wire grid polarizer (250-4000 nm) was used to measure the above-gap harmonics.

To understand the polarization properties  of HHG, a one-dimensional two-band SBEs  combined with a linear coupled excitation model is applied here. Detail of our model can be found in Appendix A. In the LCE model, one calculates one-dimensional excitation components associated with three bond directions ${\bf{e_1}}$, ${\bf{e_2}}$ and ${\bf{e_3}}$, as shown in Fig. 1(a). The laser polarization  makes an angle $\theta$ with respect to crystal's optic axis which is taken as the z-axis, and the polarization direction of harmonics are measured with respect to the laser polarization axis. Excitation along each direction is calculated using 1D two-band SBEs.  It is important to use accurate band structures and dipole matrix elements with amplitudes and phases as shown in our previous work\cite{jiang_prl}. In order to obtain the polarization distribution, a new unit vector  ${\bf{e}({\alpha})}$ pointing in the direction of ${\alpha}$ is defined.  It can be written in terms of ${\bf{e_1}}$, ${\bf{e_2}}$ as:
\begin{equation}
{\bf{e}({\alpha})}=\frac{sin({\alpha}+{\theta})}{2cos(18^{\circ})}({\bf{e_1}}-{\bf{e_2}})-\frac{cos({\alpha}+{\theta})}{2cos(72^{\circ})}({\bf{e_1}}+{\bf{e_2}})
\end{equation}
The current along ${\bf{e}({\alpha})}$, defined with respect to driving laser polarization, is written as
\begin{equation}
J({\theta},{\alpha},t)={[J_1(t){\bf{e_1}}+J_2(t){\bf{e_2}}+J_3(t){\bf{e_3}}]\cdot{\bf{e}({\alpha})}}.
\end{equation}
where ${J_j}(t)$  is the current along direction ${\bf{e_j}}$.
The intensity of the generated harmonics polarized along the direction with angle ${\alpha}$ can be calculated by projection:
\begin{equation}
I({\theta},{\alpha},{\omega})={\left|(E_1({\omega}){\bf{e_1}}+E_2({\omega}){\bf{e_2}}+E_3({\omega}){\bf{e_3}})\cdot{\bf{e}({\alpha})}\right|}^2.
\end{equation}
where ${E_j}({\omega})$  is the Fourier transform of the current along direction ${\bf{e_j}}$. In the SBEs calculation, only two bands are included.

Fig. 2 shows the comparison of experimental and theoretical polarization-resolved harmonic spectra, separated into two groups: below-gap harmonics, from H4 to H10 (a-d), and above-gap harmonics, from H11 to H17 (e-h). To illustrate orientation dependence, the theory has been scaled for each harmonic to show optimal visual agreement (each harmonic has the same scaling factor for parallel and perpendicular components). Clearly there is a good general agreement between the experimental data and the prediction from the LCE model. While some detailed features for perpendicularly-polarized above-gap harmonics are not as clearly seen because of the low signal levels, one can see that the dominant features are similar to the below-gap harmonics.

Two particularly notable features are seen from these data. When the laser polarization is parallel to the mirror plane (e.g. ${\theta} = 0^{\circ}$ or $180^{\circ}$, $\Gamma$-A axis), both even and odd harmonics have strong parallel components, but the perpendicular components disappear. When the laser polarization is perpendicular to the mirror plane, (e.g. ${\theta} = 90^{\circ}$,  $\Gamma$-M axis) all the parallel even harmonics and perpendicular odd harmonics  vanish. It is obvious that disappearance of parallel even harmonics is attributed to the reflection symmetry.  The general feature of vanishing perpendicular harmonics at specific orientation angles can also be explained by the symmetry properties of the system, as shown below.

By inserting eq. (1)   into eq.(2), the latter can be expressed as

\small{
\begin{align}
J({\theta},{\alpha},t)=
&A({\theta},{\alpha})({J_1}(t)+cos(144^{\circ}){J_2}(t)+cos(108^{\circ}){J_3}(t))+\nonumber\\
&B({\theta},{\alpha})(cos(144^{\circ}){J_1}(t)+{J_2}(t)+cos(108^{\circ}){J_3}(t)),
\end{align}
}
where
\small{
\begin{align}
A({\theta},{\alpha})=\left(\frac{sin({\alpha}+{\theta})}{2cos(18^{\circ})}-\frac{cos({\alpha}+{\theta})}{2cos(72^{\circ})}\right),\nonumber\\
B({\theta},{\alpha})=\left(\frac{-sin({\alpha}+{\theta})}{2cos(18^{\circ})}-\frac{cos({\alpha}+{\theta})}{2cos(72^{\circ})}\right)\nonumber
\end{align}
}
When ${\alpha}$ is set to $90^{\circ}$, we can obtain the perpendicular current. For the  two special cases, e.g., ${\theta}$ = $0^{\circ}$ and $90^{\circ}$ ,
$J({\theta}=0^{\circ},{\alpha}=90^{\circ},t){\propto} {J_1}(t)-{J_2}(t)$, and $J({\theta}=90^{\circ},{\alpha}=90^{\circ},t){\propto} {J_1}(t)+{J_2}(t)$, respectively.
Using the strong field approximation (SFA),

\small{
\begin{align}
&J({\theta}=0^{\circ},{\alpha}=90^{\circ},t){\propto}\nonumber\\
&-i\omega\int\limits_{BZ}d{k}\int_{-\infty }^{t}dt'D_{cv,e_{1}}^{*}(k+A_{e_{1}}(t))F_{e_{1}}(t')D_{cv,e_{1}}(k+A_{e_{1}}(t'))\nonumber\\
&e^{-iS(k,t,t')}\nonumber\\
&+i\omega\int\limits_{BZ}d{k}\int_{-\infty }^{t}dt'D_{cv,e_{2}}^{*}(k+A_{e_{2}}(t))F_{e_{2}}(t')D_{cv,e_{2}}(k+A_{e_{2}}(t'))\nonumber\\
&e^{-iS(k,t,t')}+c.c.
\end{align}
}
where $S(k,t,t')$ is the standard action in the SFA theory and $D_{cv,e_i}$ is the transition dipole moment. For ${\theta}=0^{\circ}$,$A_{e_{1}}=A_{e_{2}}$,$F_{e_{1}}=F_{e_{2}}$,$D_{cv,e_{1}}=D_{cv,e_{2}}$, which leads to $J({\theta}=0^{\circ},{\alpha}=90^{\circ},t)=0$ .
In turn, this leads to   no perpendicular harmonic signal, for both even and odd harmonics,  for ${\theta}=0^{\circ}$, in agreement with experimental observation and the prediction of the LCE model.  Similarly, we obtain

\small{
\begin{align}
&J({\theta}=90^{\circ},{\alpha}=90^{\circ},t){\propto}\nonumber\\
&-i\omega\int\limits_{BZ}d{k}\int_{-\infty }^{t}dt'D_{cv,e_{1}}^{*}(k+A_{e_{1}}(t))F_{e_{1}}(t')D_{cv,e_{1}}(k+A_{e_{1}}(t'))\nonumber\\
&\times e^{-iS(k,t,t')}\nonumber\\
&-i\omega\int\limits_{BZ}d{k}\int_{-\infty }^{t}dt'D_{cv,e_{2}}^{*}(k+A_{e_{2}}(t))F_{e_{2}}(t')D_{cv,e_{2}}(k+A_{e_{2}}(t'))\nonumber\\
&\times e^{-iS(k,t,t')}+c.c.
\end{align}
}
For ${\theta}=90^{\circ}$, $A_{e_{1}}=-A_{e_{2}}$, $F_{e_{1}}=-F_{e_{2}}$,$D_{cv,e_{1}}=D_{cv,e_{2}}$. This means that the quantum orbit at $(k_s,t'_s,t_s)$ and at  $(k_s,t'_s+T/2,t_s+T/2)$ are always equal. This symmetry leads to the existence of pure even harmonics in the perpendicular direction, which is consistent with the prediction of the LCE model as well as the experimental observation. When the polarization of the driving laser is not parallel or perpendicular to the optic axis, both even and odd harmonics would appear in general.

To show these symmetry-imposed features more clearly, we display the polarization-resolved orientation-dependent intensities of harmonics H5 and H6 in Fig. 3. Similar plots for H11 and H12 are shown in the Appendix B. For the parallel component of  H5, the experiment shows additional peaks at about 45 and 135 degrees. These peaks are absent in the theory. Similar weaker features also appear at H7 in the experiment, but not in above-gap harmonics. This difference is likely due to the birefringence properties\cite{xia_optexpress} of bulk ZnO crystals, as discussed in Appendix C. As the harmonics are generated close to the exit plane of the crystal, they are sensitive to changes in the driving laser polarization, which becomes elliptical for crystal orientation angles close to $45^{\circ}$ and $135^{\circ}$. For other polarization components, all the notable features agree very well between experiment and simulation.
\begin{figure}
\centering
    \includegraphics[width=3.5in,angle=0]{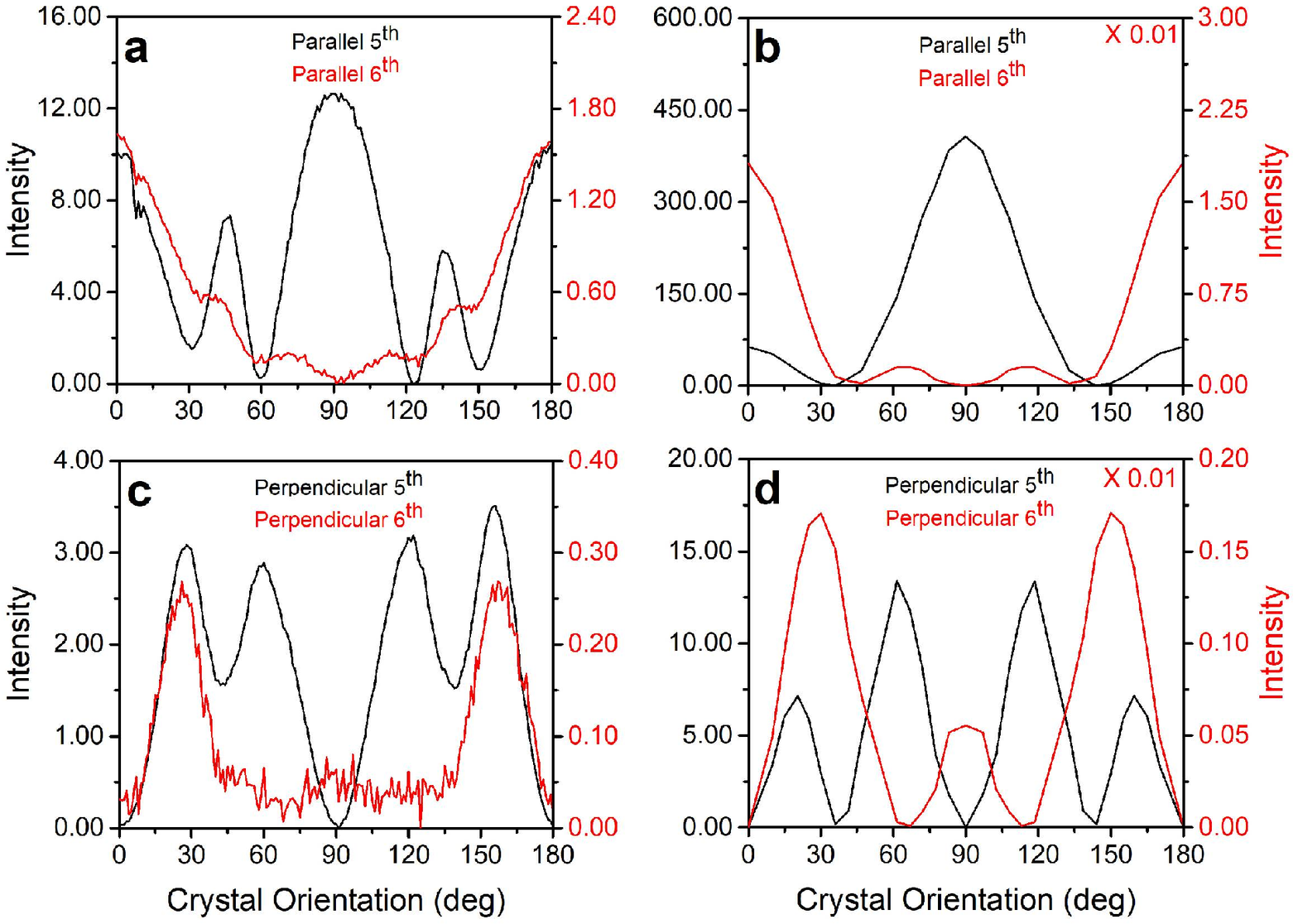}
    \caption* \justifying{Fig. 3. The top frames are for the parallel components and the bottom frames are for the perpendicular components. The left column is from experiment and the right column is from simulation. For parallel components, the additional peaks observed (a) at $45^{\circ}$ and $135^{\circ}$  but not seen in simulation (b) are attributed to birefringent effect, see text. For the perpendicular polarization, the positions of the maxima and minima in the experiment and simulation are identical, but the relative magnitude of the peaks differs.  }\label{Fig_3}\justifying
\end{figure}

For a particular orientation of the crystal, we also determine the polarization states of the emitted harmonics. Specifically, we characterize the harmonic polarization for driving laser polarizations both parallel and perpendicular to the crystal c-axis. The results from experiment and theory for below-gap harmonics are shown in Fig. 4, while above-gap harmonics are shown in Appendix B. We find generally good agreement between theory and simulation for each harmonic. Along the c-axis (upper frames), the polarization states of even and odd harmonics are quite similar since both even and odd harmonics have large parallel components, while the perpendicular components are negligible. Thus the signal decreases monotonically as the polarizer is rotated from ${\alpha} = 0^{\circ}$ ($180^{\circ}$) to $90^{\circ}$. When the laser polarization is perpendicular to the c-axis (lower frames), the perpendicular polarization components of odd harmonics vanish, while the parallel component is large. Thus  the  odd harmonics vanish at ${\alpha} = 90^{\circ}$. On the other hand, for even harmonics, the behavior is reversed: the parallel component vanishes, while the perpendicular component is nonzero. Thus for polarizer at ${\alpha} = 90^{\circ}$, the signal is large, and it decreases as the angle moves away from it.

\begin{figure}[!htbp]
\centering
    \includegraphics[width=3.5in,angle=0]{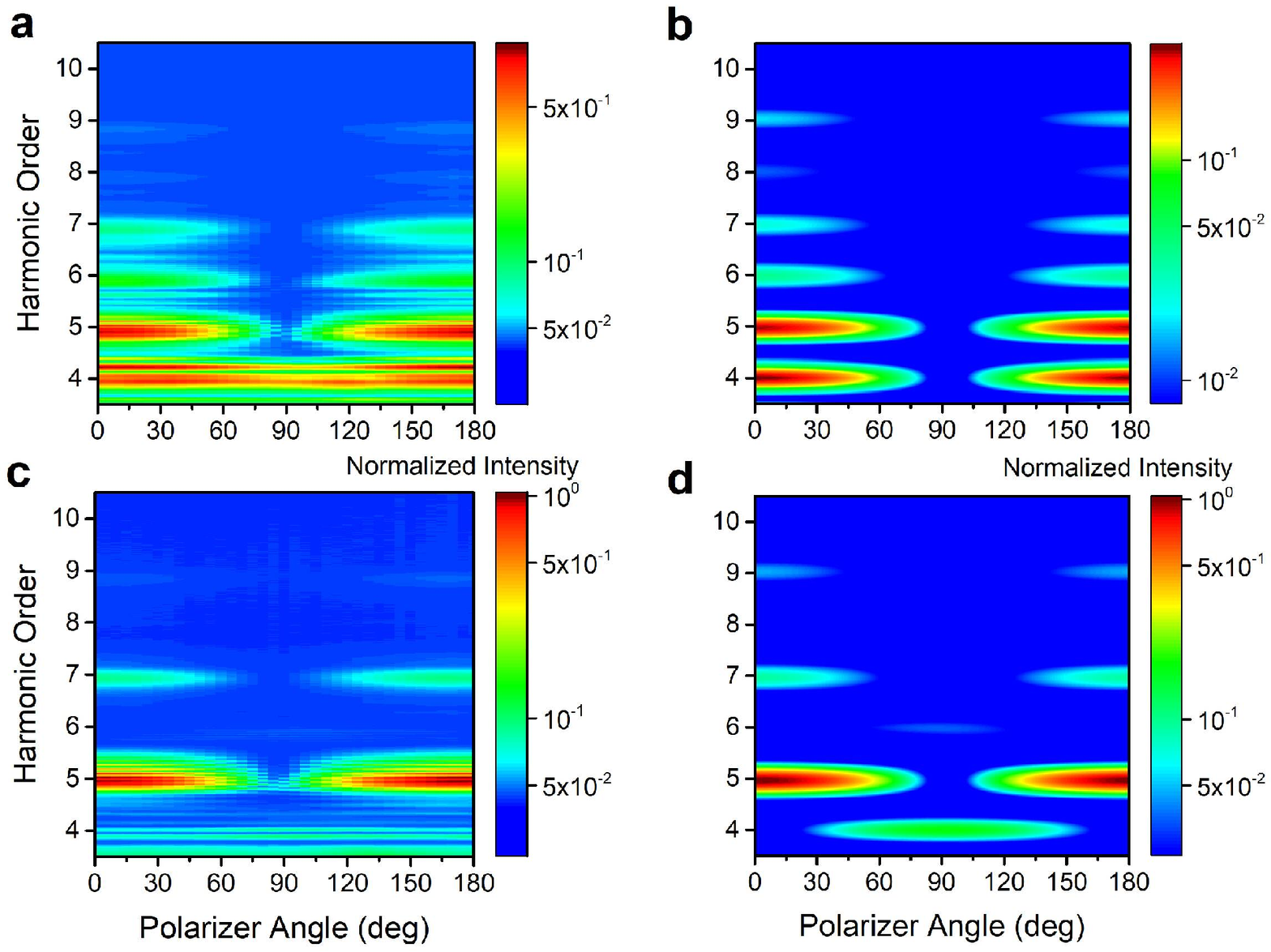}
    \caption* \justifying{Fig. 4. The driving laser is polarized parallel (top frames) or perpendicular (bottom frames) to the optic axis of the crystal. The left column is from experiment and the right column is from simulation. Each harmonic in the simulation is normalized independently for better comparison with experiment. The signals are obtained in experiment by rotating the polarizer.}\label{Fig_4}\justifying
\end{figure}
\section{Discussion on the mechanisms for perpendicularly polarized harmonics}
The above discussion on the polarizations of harmonics concentrated only on ZnO, calculated using the 1D two-band SBEs together with the LCE model. In this case, the theory explains the experiments quite well. However, the arguments are mostly based on the symmetry of the crystal. Thus we wish to determine whether the observed behavior can be generalized to other systems. By perusing experimental polarization data for different solids, we find that most data show similarities to ZnO, provided that the symmetry properties of the material are considered. We first consider the results of Ref. 22, in which THz pulses are used to generate harmonics from GaSe. All the harmonics reported in this experiment are below the band gap. Their Figs. 1c and 1d are similar to our Fig. 4. Their simulation, which is based on solving the 1D SBEs with three valence bands and two conduction bands together with the LCE model, indeed also explained the observed polarization data. Polarization angle dependence was also reported in Ref. 24 for $\alpha$-quartz. Their  Fig. 2 graphs, though plotted differently, agree qualitatively with our Fig. 4. The perpendicular harmonics in Ref. 24 were interpreted using the intra-band Berry curvature, and thus the underlying theory appears to be different from the one used here and in Ref. 22. Additional experimental data can be found in Ref. 23 for MoS$_2$ and in Ref. 25 for GaSe. All of these experimental results show general features that: 1) parallel even harmonics and perpendicular odd harmonics vanish when the laser polarization is perpendicular to the mirror plane; 2) perpendicular even and odd harmonics vanish when the laser polarization is parallel to the mirror plane.

From these  results, one may conclude that many of the features of polarization-dependent HHG spectra are governed by the crystal symmetry, irrespective of the excitation ``mechanisms''. we will show that all these mechanisms, including Berry curvature, interband excitation and band curvature, are consistent with each other. Because wurtzite ZnO has broken symmetry, we have ${{\bf D}_{m,n}(-k_x,-k_y,-k_z)={\bf D}_{m,n}^*(k_x,k_y,k_z)}$ for the transition dipole in a specific gauge. Once the system has mirror symmetry, say on the $(x=0,y,z)$ plane, we can get $u_{m,-k_x,k_y,k_z}(x,y,z) = \pm u_{m,k_x,k_y,k_z}(-x,y,z)$ where $u_{m,{\bf k}}(\bf r)$ is the periodic part of the Bloch eigenfunction. When   $``+" (-) $ is for band $m$ and$``-" (+)$ is for band $n$, we say they have opposite ``parities". Otherwise, we say they have same ``partities". If band $m$ and band $n$ have the same ``parities", we find:
\begin{align}
D_{m,n}^x(-k_x,k_y,k_z)=\frac{ip_{m,n}^x(-k_x,k_y,k_z)}{\Delta E_{m,n}} =-D_{m,n}^x(k_x,k_y,k_z)
\end{align}
\begin{align}
D_{m,n}^z(-k_x,k_y,k_z)=\frac{ip_{m,n}^z(-k_x,k_y,k_z)}{\Delta E_{m,n}} =D_{m,n}^z(k_x,k_y,k_z)
\end{align}
where $p_{m,n}^{\alpha} ({\alpha} =x, y, z)$ is the momentum matrix, ${\Delta E_{m,n}}$ is the energy difference between band $m$ and band $n$.
Based on these two equations, along the $\Gamma - $A direction where $k_x = k_y = 0$, we find that $D^x_{m,n}(k_z)$ is zero and therefore the perpendicular components of even and odd harmonics both disappear. Along the $\Gamma - $M direction, $k_y = k_z = 0$, and thus $D^x_{m,n}(k_x)$ and $D^z_{m,n}(k_x)$ have opposite reflection ``parities'', e.g. $D^x_{m,n}(k_x)$ is an odd function and $D^z_{m,n}(k_x)$ is an even function with respect to $k_x$. This leads to odd and even harmonics with purely parallel and perpendicular polarizations, respectively. If the eigenfunctions of valence band and conduction band have opposite ``parities" we come to get a similar conclusion but with z and x components interchanged. These results are consistent with Fig.~3. The conclusion above is also consistent with the  Berry curvature mechanism.  From Kubo formula \cite{Yao_curvature}, Berry curvature is given by
\begin{equation}
{\Omega}_n^y=-\sum\limits_{m\ne n}\frac{2Im(p_{n,m}^x p_{n,m}^z)}{\Delta{ E_{m,n}^2}}
\end{equation}
When the driving laser is along $\Gamma - $A, $p_{n,m}^x (k_z)= 0$ or  $p_{n,m}^z(k_z) = 0$ , ${\Omega}_n^y(k_z) = 0$ and no perpendicular harmonics are induced by Berry curvature. When the driving laser is along $\Gamma - $M,  $p_{n,m}^x(k_x)$ and  $p_{n,m}^z(k_x)$ are nonzero and have opposite ``parities'', resulting in Berry curvature being an odd function of $k_x$. This would lead to purely perpendicular even harmonics according to the Berry Curvature mechanism \cite{shambhu_mos2}.  The simulated Berry curvature ${\Omega}_n^y$ of the first conduction bands for $\Gamma - $A and $\Gamma - $M directions in Fig. 5.

From the previous work\cite{GaSe-PRL}, perpendicular harmonics can also come from band curvature. In the second-order nondegenerate perturbation theory, the band energy around $\bf k_0$ can be expressed as \cite{S. W. Koch's book}
\begin{align}
E_n({\bf k})&={\bf k}^{2}/2+{\bf k}\cdot {\bf p}_{nn}({\bf k}_0)+E_n({{\bf k}_0})\nonumber\\
&+\sum\limits_{\lambda\ne n}\frac{({\bf k}\cdot {{\bf p}_{n,\lambda}}({\bf k_0}))({\bf k}\cdot {{\bf p}_{\lambda,n}}({\bf k_0}))}{ E_n(\bf k_0)-E_{\lambda}(\bf k_0)}
\end{align}
Band curvature can be expressed as:
\begin{align}
\frac{\partial ^{2}E_n{(\bf k)}}{\partial k_i \partial k_j}=\frac{1}{m_{n,ij}}={\delta_{ij}}+\sum\limits_{\lambda\ne n}\frac{2p_{n\lambda,i}({\bf k_0})p_{\lambda n,j}({\bf k_0})} { E_n(\bf k_0)-E_{\lambda}(\bf k_0)}
\end{align}
where ${m_{n,ij}}$ is the effective mass tensor. Thus, similar to Berry curvature in Kubo's formula, the band curvature can also be expanded by inter-band momentum matrix. Because the band curvature induced electric field ${\bf E}^{bc}(t)\propto \frac{F(t)}{m_{n,ij}}$, it is easy to prove that $E^{bc}_{\perp}(t)=E^{bc}_{\perp}(t+T/2)$ and $E^{bc}_{||}(t)=-E^{bc}_{||}(t+T/2)$ if the driving laser's polarization is perpendicular to miror plane. It means that all these mechanisms, including Berry curvature, interband excitation and band curvature, are consistent with each other, and polarizations of harmonics are governed by crystal symmetry, not by one of the ``mechanisms". Thus, when such symmetry imposed constraints are not satisfied in simulations or in experiments, due care must be exercised.

\begin{figure}[!htbp]
\centering
    \includegraphics[width=3.5in,angle=0]{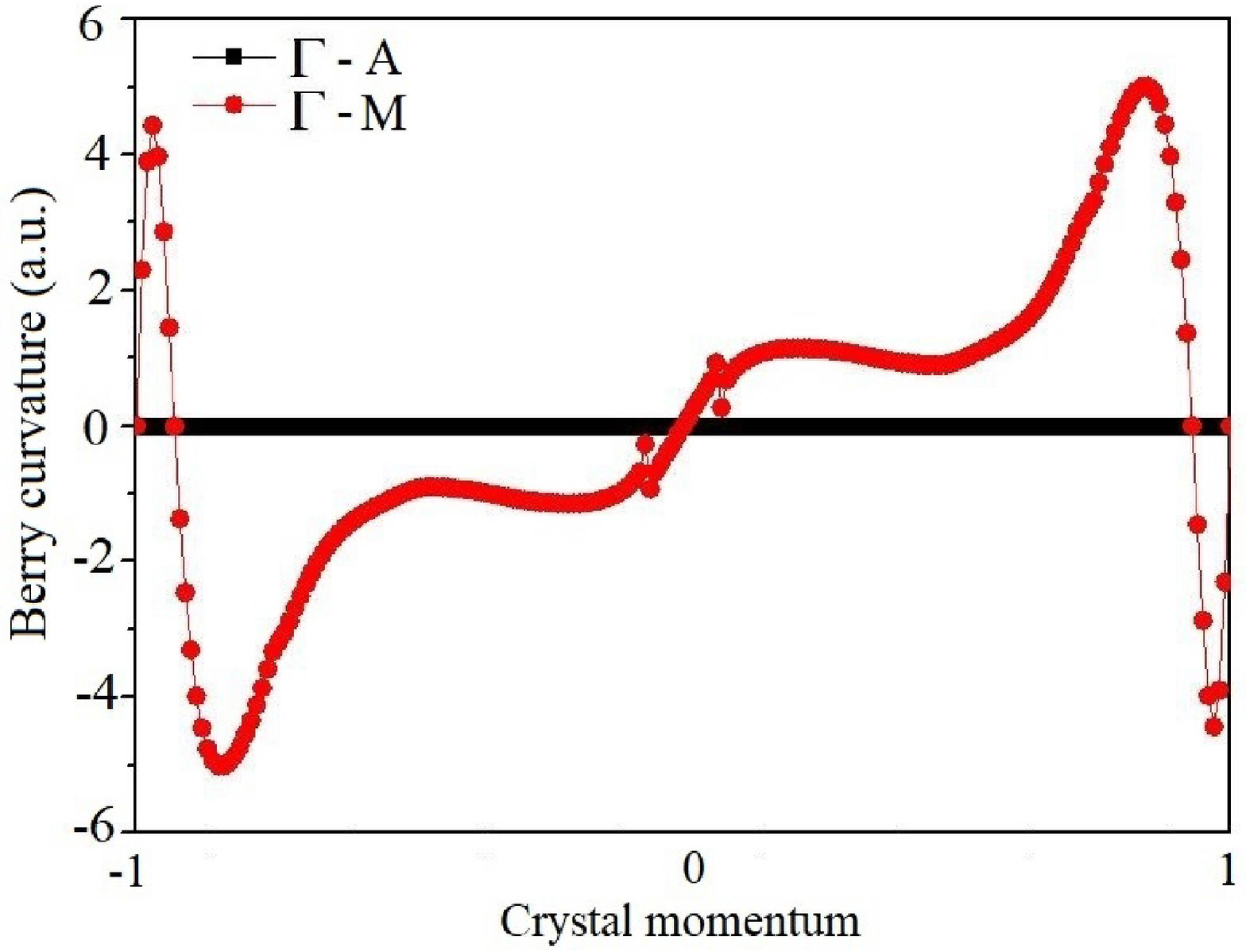}
    \caption* \justifying{Fig. 5. Berry curvature of the first conduction bands for  $\Gamma - $A and $\Gamma - $M. For both of these directions, the number of k points is set to be 200, and 48 bands with 18 valence bands are included}\justifying
\end{figure}

\section{Polarization ellipse of harmonics}
From LCE model, it is clear that the polarization states of HHG are dependent on the bond structure of the crystal. Because the bonds inside crystals are fixed naturely, it is a potential way to tune the ellipse states of the HHG by changing the structures though strain or selecting crystals with specific structures. Here we show the ellipticity of HHG from wurtzite ZnO briefly. There is a phase angle in the electric fields between the two polarization components. Such phase angle can be determined experimentally in principle, but has not been done so far for harmonics from solids. If this phase angle ${\delta}$ is determined, then the polarization is fully described by an ellipse, see Fig. 1, where the ellipse is characterized by an orientation angle $\phi$ with respect to the polarization of the driving laser and the ellipticity  $\epsilon$.  These two experimentally determined quantities $\phi$ and $\epsilon$ are related to parallel and perpendicular harmonics strength and the relative phase angle ${\delta}$ through\cite{ ATLpol}
\begin{equation}
tan(2{\phi})=tan(2{\gamma})cos({\delta})
\end{equation}
\begin{equation}
sin(2{\chi})=sin(2{\gamma})sin({\delta})
\end{equation}
where $\chi$ and $\gamma$ are defined by  $\epsilon$ = $tan({\chi})$ and $tan({\gamma})=\sqrt{\frac{I_{\perp}}{I_{\parallel}}}$.

Here we calculate the orientation angle and ellipticity from the LCE model for the above-gap harmonics  (right column of Fig. 6), at crystal orientation angles ${\theta}$ = $0^{\circ}$, $20^{\circ}$, $72^{\circ}$ and $90^{\circ}$, respectively.
Although the relative phase angle ${\delta}$ can't be determined experimentally at present, the upper bound of ellipticity $\epsilon$ and orientation angle $\phi$ can be extracted by Malus' law\cite{Hecht} as shown in the left column of Fig. S8. At $0^{\circ}$ and $90^{\circ}$, the harmonics are linearly polarized and the calculations are consistent with the expected results. At $20^{\circ}$ and $72^{\circ}$, small ellipticity of about 0.1 and small orientation angles of about $10^{\circ}$  to $20^{\circ}$ have been found. Their small values  is consistent with the smaller perpendicular harmonic components seen in ZnO.

\begin{figure}[!htbp]
\centering
    \includegraphics[width=3.5in,angle=0]{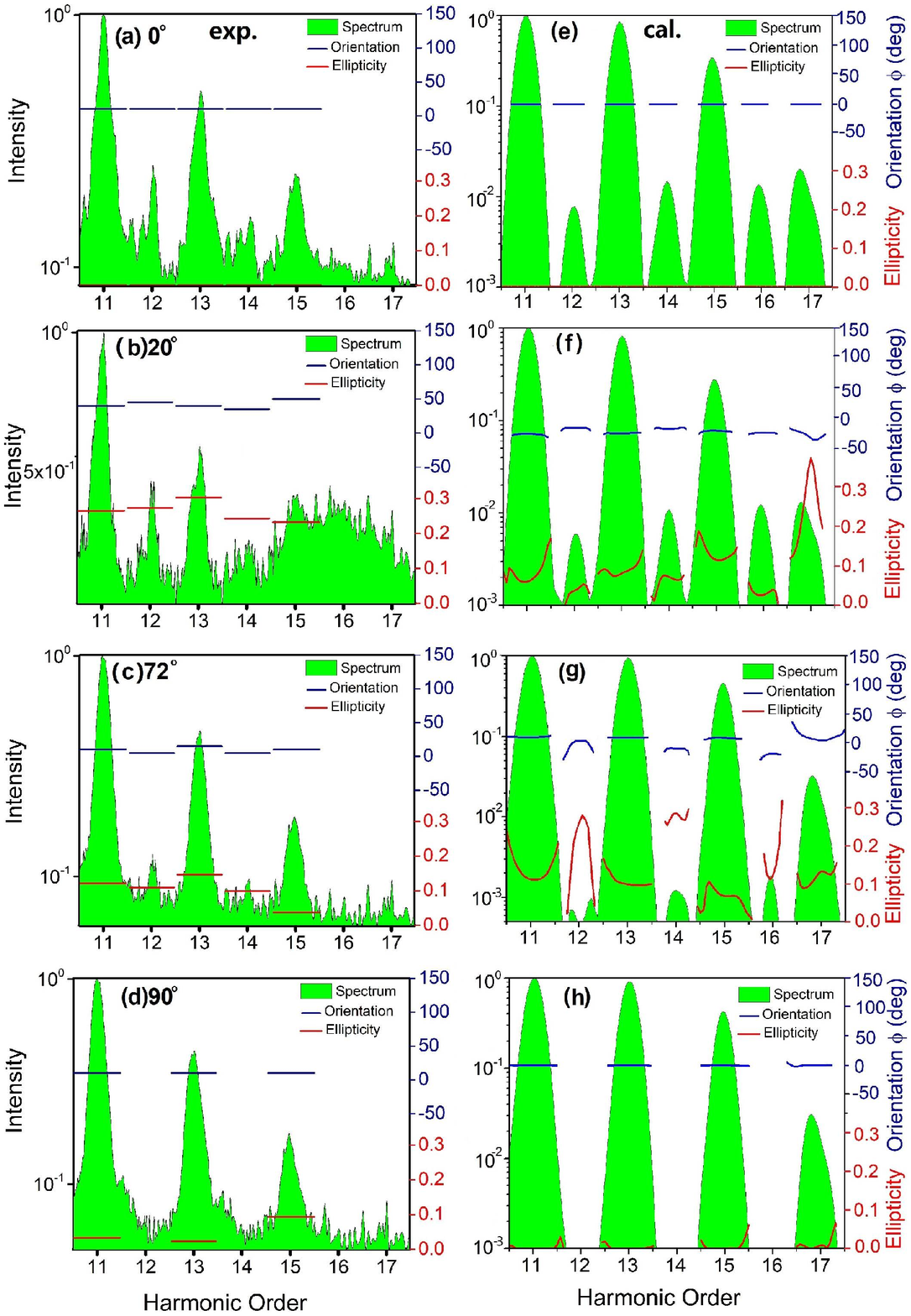}
    \caption* \justifying{ Fig. 6. Left column shows experimentally extracted upper bound of  ellipticity and orientation angle of the polarization ellipse of the harmonic signals.  Right column shows theoretically predicted ellipticity and orientation angle of the polarization ellipse.   (a) (e) ${\theta}=0^{\circ}$,(b) (f) ${\theta}=20^{\circ}$,(c) (g) ${\theta}=72^{\circ}$,(d) (h) ${\theta}=90^{\circ}$, respectively. The green area represents the harmonic spectra. Blue lines are the orientation angles and red lines are the ellipticities.}\justifying
\end{figure}

\section{Summary}
The 1D two-band SBEs model employed in our calculation is still not complete. Since the phase of the transition dipole cannot be obtained from the commercial \emph{ab-initio} codes, our calculations rely upon the tight-binding model to calculate the phase\cite{jiang_prl}. The yields for even harmonics from the present theory  appear to drop too quickly with the increase of the harmonic order, but the orientation dependence of the polarization of the harmonics agrees well with experiments.  Until now, Berry curvature \cite{shambhu_mos2,berrycurvature-TTLu}, band curvature \cite{GaSe-PRL} and inter-band polarization \cite{Langer_naturephoton2017} have been used to explain the perpendicular and parallel components of harmonic spectra. However, each alone is only an approximation to a complete theory that is yet to be developed. For now, a 1D SBEs model that accounts for Berry phase would be a first step toward such a theory.

In summary, we have demonstrated that polarization properties of  high-order harmonics generated in ZnO  can be explained using 1D two-band SBEs combined with the linearly coupled excitation (LCE) model. By comparing the existent experimental data for different systems, we emphasize that the polarization properties of high-order harmonics in solids are governed largely by the symmetry properties of the crystal. These results show that polarization of harmonics generated in a crystal provide a powerful tools for probing the spatial symmetry properties of a crystal. Using femtosecond pulses to generate harmonics, measurement of harmonic polarization states offers the opportunity to probe structural changes in a crystal with unprecedented temporal resolution.
\section*{Acknowledgements}
SJ and SG equally contributed to this work. CDL was supported in part by the Chemical Sciences, Geosciences, and Biosciences Division, Office of Basic Energy Sciences, Office of Science, U.S. Department of Energy, under Grant No. DE-FG02-86ER13491. The work done at UCF was supported by the Air Force Office of Scientific Research under Award No. FA9550-16-1-0149 and by the National Science Foundation under Grant No. 1806135. SJ's new adress is State Key Laboratory of Precision Spectroscopy, East China Normal University, Shanghai 200062, China.

\section*{Appendix A \\ SBEs combined with LCE model}
In the model of linear combination of excitations (LCE), one calculates one-dimensional Semiconductor Bloch Equations (SBEs) associated with three bond directions ${\bf{e_1}}$, ${\bf{e_2}}$ and ${\bf{e_3}}$, as shown in Fig. 1(a). For each direction, we use two-band SBEs model which can be written as \cite{jiang_pra}

\small{
\begin{align}
\frac{{\partial \rho_{cv}({\bf{k}},t)}}{{\partial t}}=\left(-iE_g({\bf{k}})-\frac{1}{T_2} \right)\rho_{cv}({\bf{k}},t)
+{\bf{F}}(t)\cdot\nabla_{\bf{k}}\rho_{cv}({\bf{k}},t)\nonumber\\
+i\left[{\rho_c({\bf{k}},t)-\rho_v({\bf{k}},t)}\right]{\bf{F}}(t)\cdot{\bf{D}}_{cv}({\bf{k}}),\tag{A1}
\end{align}
\begin{align}
\frac{{\partial \rho_v({\bf{k}},t)}}{\partial t}=-2{\mathop{\rm Im}\nolimits}\left\{{\bf{F}}(t)\cdot{\bf{D}}_{cv}({\bf{k}})\rho_{cv}({\bf{k}},t)\right\}
+{\bf{F}}(t)\cdot\nabla_{\bf{k}}\rho_v({\bf{k}},t),\tag{A2}
\end{align}
\begin{align}
\frac{\partial \rho_c({\bf{k}},t)}{\partial t}=2{\mathop{\rm Im}\nolimits}\left\{{\bf{F}}(t)\cdot{\bf{D}}_{cv}({\bf{k}})\rho_{cv}({\bf{k}},t)\right\}
+{\bf{F}}(t)\cdot\nabla_{\bf{k}}\rho_c({\bf{k}},t).\tag{A3}
\end{align}
}Here $\rho_{cv}({\bf{k}},t)$  is the micropolarization between the conduction band and the valence band. $E_{g}({\bf{k}})=E_{c}({\bf{k}})-E_{v}({\bf{k}})$ is the energy difference between the two bands,  $\rho_{c(v)}({\bf{k}},t)$ is the electron density in the conduction (valence) band. A k-dependent bell-shaped form of the dephasing time  $T_2(k)=1+ 1/[1+ exp( 100|k|-5)$ is used in our calculation. From the SBEs, we can calculate the current  ${J_j}(t)$ induced by the laser field along direction ${\bf{e_j}}$ by:
\small{
\begin{align}
{\bf{J}}_{i}(t)&=\sum\limits_{\lambda=c,v}\int\limits_{BZ}{\bf{v}}_\lambda({\bf{k}})\rho_\lambda({\bf{k}},t)d{\bf{k}}
+\frac{\partial}{{\partial t}}\int\limits_{BZ}{\bf{D}}_{cv}({\bf{k}})\rho_{cv}({\bf{k}},t)d{\bf{k}}\nonumber\\
&+c.c,\tag{A4}
\end{align}
}
Using the specific coordinate system and angles as defined in Fig. 1(a) of the main text, the current along a direction with an angle ${\alpha}$ is obtained by projection:
\small{
\begin{align}
J({\theta},{\alpha},t)={[J_1({t}){\bf{e_1}}+J_2(t){\bf{e_2}}+J_3(t){\bf{e_3}}]\cdot{\bf{e}({\alpha})}}.\tag{A5}
\end{align}
}
Here
${\theta}$ is the polarization angle of the driving laser, and ${\alpha}$ is defined with respect to the laser's polarization.  A detailed description of how we obtain the band structure  and complex dipole moment are given in our previous paper \cite{jiang_prl}. High-order harmonics signal can be calculated by finding the Fourier-transform of the current, i.e.

\small{
\begin{align}\label{eq_SHHG}
S_{\mathrm{HHG}}({\theta},{\alpha},{\omega})\propto\left|\int_{-\infty}^{\infty}\left[{\bf{J({\theta},{\alpha},t)}}(t)\right]e^{i\omega t}dt\right|^2.\tag{S6}
\end{align}
\section*{Appendix B \\ Orientation-dependent and polarization distribution of above-gap harmonics}
In Fig. 2 and Fig. 3 of the main text, it was shown that the parallel component of the below-gap harmonic H5 shows maxima at the orientation angles of $45^{\circ}$ and $135^{\circ}$ in the experimental data but not  in the prediction of the LCE model. From Fig. 2(a) of the main text, such maxima also appear weakly for H7. For above-gap harmonics, the signals are weaker but these peaks are absent. In Fig. A1, we present the line-out of the orientation-dependent harmonics H11 and H12, for both the parallel and perpendicular components. In Fig. A1, clearly there are no peaks near $45^{\circ}$ and $135^{\circ}$, like those seen in H5. There is a good agreement in the parallel components in H11 and H12 between the experiment and the LCE model. For the weaker perpendicular components, the agreement is also very good despite that the noise in the experiment becomes much more pronounced. Similar to Fig.4 for below-gap harmonics, in Fig. A2 we show the polarization distributions of harmonics at two crystal orientations, for the polarization of the driving laser parallel and perpendicular to the crystal axis. The general polarization distributions for above- and below-gap harmonics are essentially identical.
\begin{figure}[!htbp]
    \includegraphics[width=3.5in,angle=0]{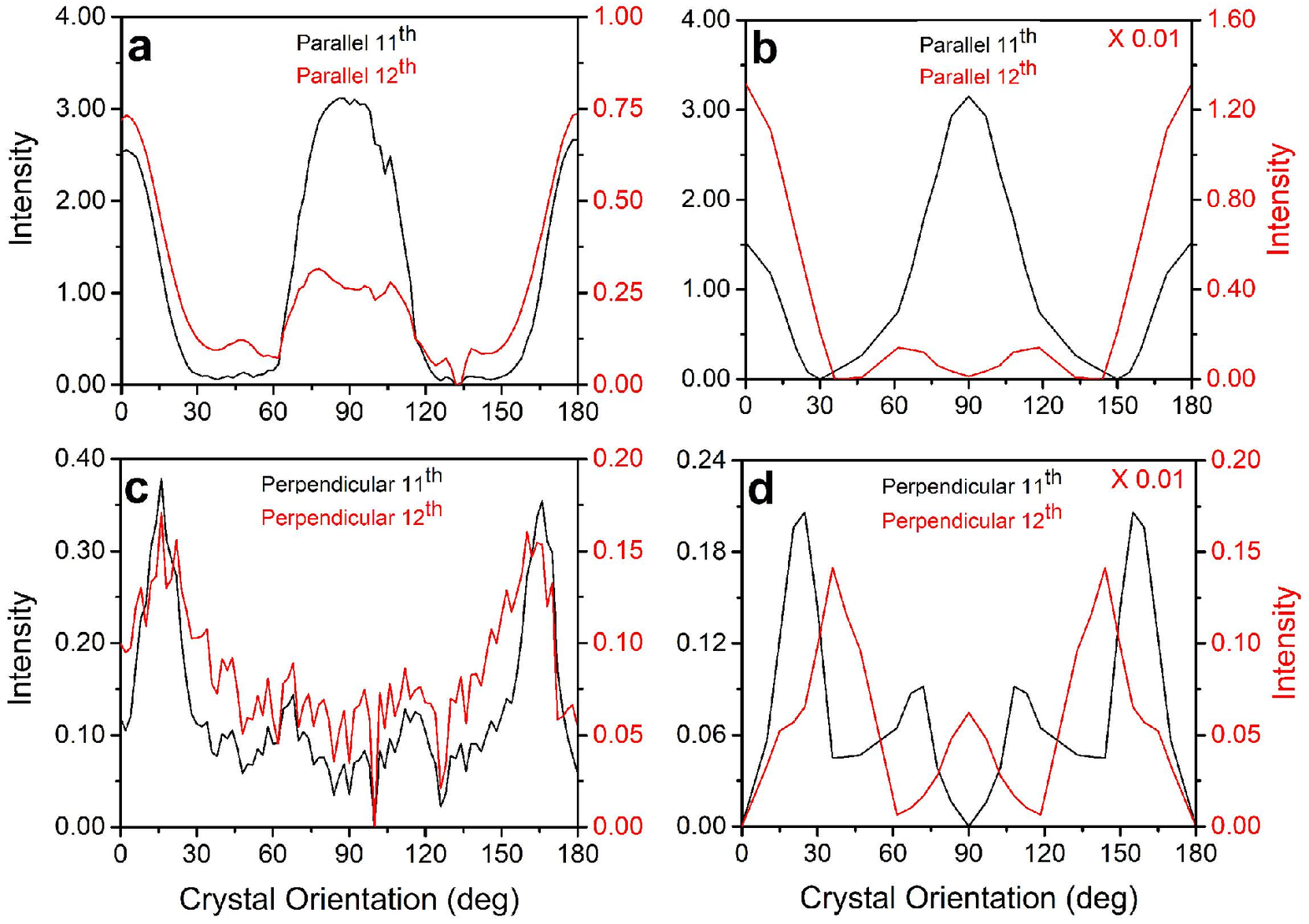}
    \caption* \justifying{ Fig. A1. Orientation-dependent intensity of H11 and H12, for parallel (a), (b) and perpendicular (c), (d) components, respectively. (a), (c) are the experimental data and (b), (d) are from the simulations.}\justifying
\end{figure}

\begin{figure}[!htbp]
\centering
    \includegraphics[width=3.5in,angle=0]{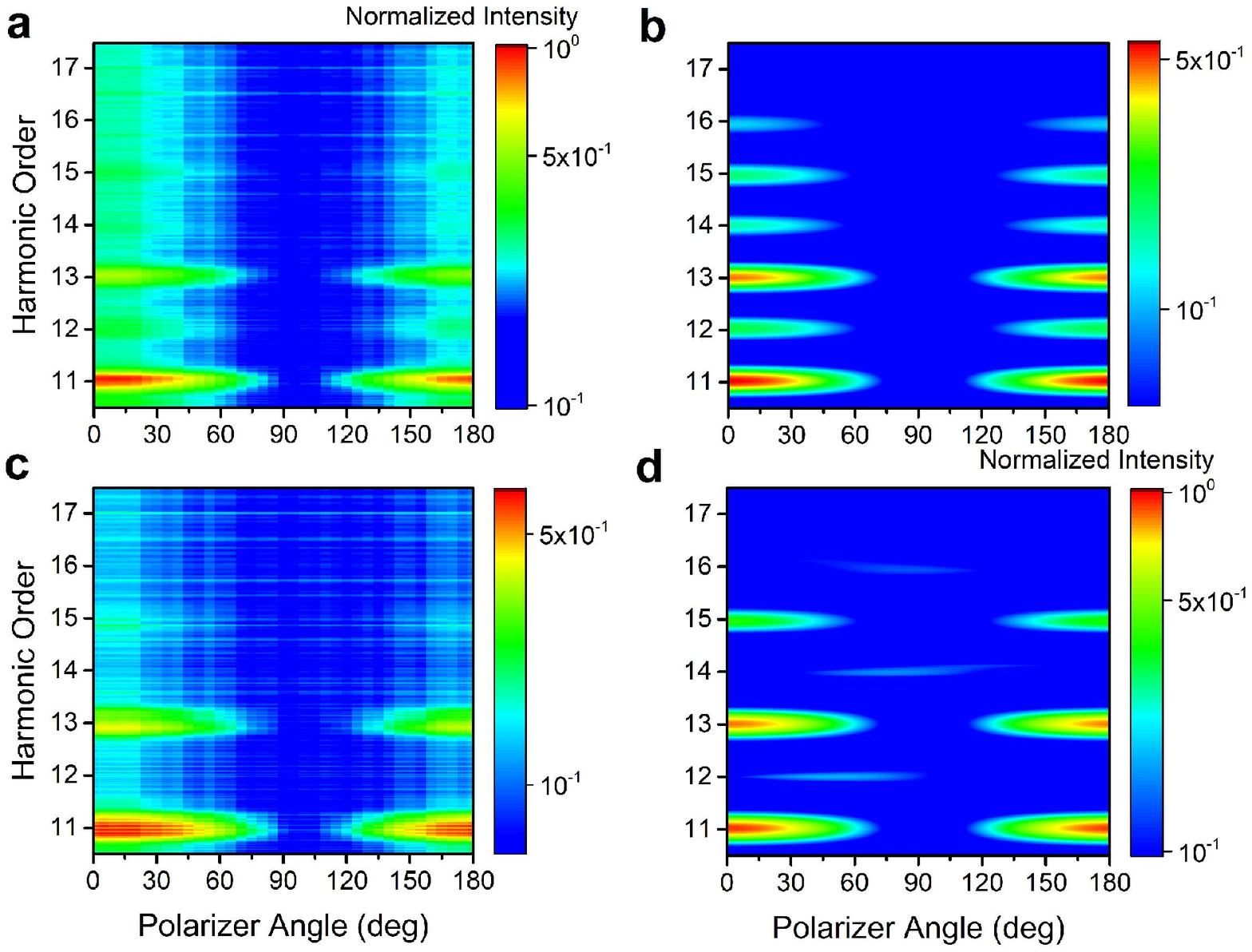}
    \caption* \justifying{Fig. A2.  Polarization distribution of above-gap harmonics when the driving laser is polarized parallel [(a), (b)] and perpendicular [(c) , (d)] to the optic axis. (a), (c) are detected experimentally by rotating the polarizer, and (b), (d) are the corresponding calculated results.}\justifying
\end{figure}
\section*{Appendix C \\ Effects of ZnO Birefringence}
The birefringence properties of ZnO may lead to undesired changes in the polarization states of both the mid-IR driving laser pulses and the generated high-order harmonics. Ideally, harmonics could be generated in a thin ZnO film; however, monocrystalline films of a-plane ZnO are not available to us at this time. Due to the experimental geometry, however, harmonics are primarily generated within a small volume close to the exit plane of the crystal, and therefore the dominant role of birefringence will be to change the driving laser polarization. We have measured the ellipticity of the driving laser after propagation through the crystal under the experimental conditions. The result is shown below in Fig. A3. We find that the driving laser ellipticity increases from 0 to 0.8 as the orientation angle $\theta$ is increased from 0$^{\circ}$ to 45$^{\circ}$, and then decreases back to 0 at an orientation angle of 90$^{\circ}$. This behavior is in good agreement with the known birefringence properties of ZnO, as shown in the figure, indicating that the effects of nonlinear propagation on the polarization are small.
\begin{figure}[!htbp]
\centering
    \includegraphics[width=3.5in,angle=0]{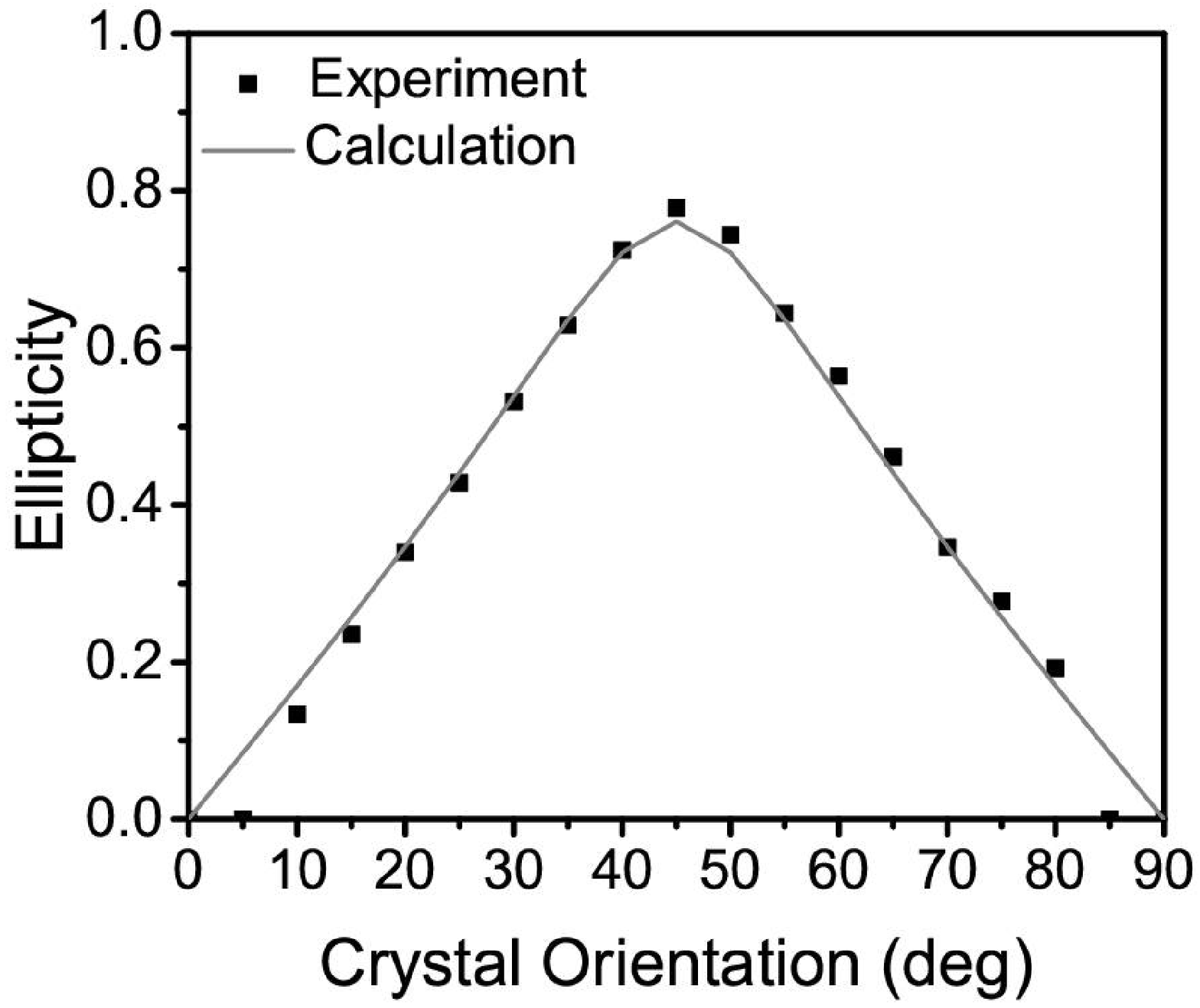}
    \caption* \justifying{Fig. A3. Orientation dependent ellipticity of driving laser at the crystal exit plane. The gray solid line shows the calculated ellipticity based on the known birefringence of ZnO and the crystal thickness of 0.3 mm.}\justifying
\end{figure}
\begin{figure}[htbp]
    \includegraphics[width=3.5in,angle=0]{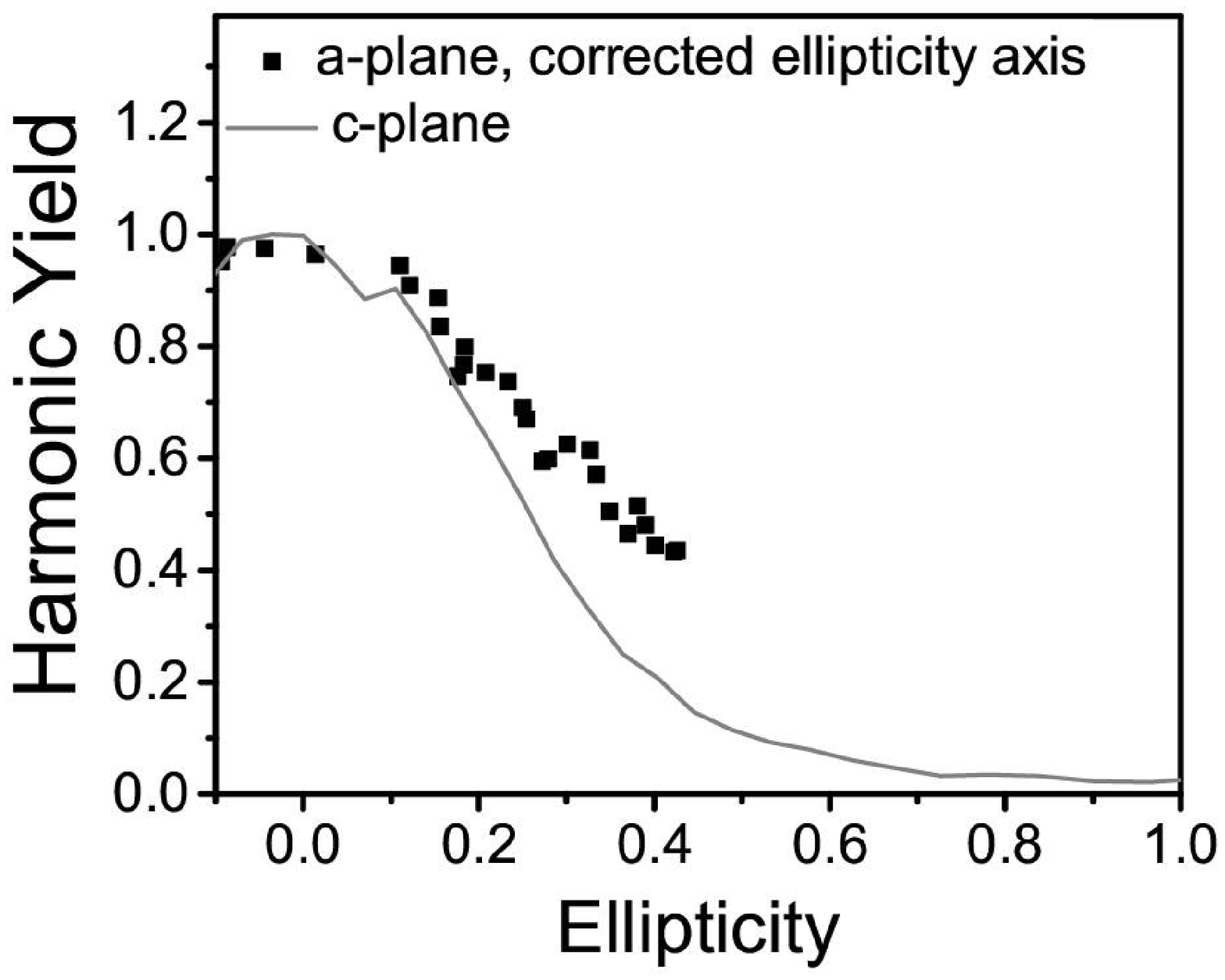}
    \caption* \justifying{Fig. A4. Dependence of 11th harmonic yield on the ellipticity of driving laser, for laser polarization in the a-plane (dots) and c-plane (solid gray line) of ZnO. For the a-plane measurements, the ellipticity of the driving laser after propagation through the crystal was measured experimentally, while for the c-plane, we confirmed that the polarization state was not changed during propagation.}\justifying
\end{figure}
It is well known that driving laser ellipticity affects high harmonic generation in solids \cite{ZnO-2011}. However, measurements of ellipticity dependence in the a-plane of ZnO have not been reported. Fig. A4 shows the ellipticity dependence of the 11th harmonic generated for laser polarization in the a-plane of ZnO. The laser ellipticity was set using a combination of a half-wave plate and a quarter-wave plate, and the data were corrected for propagation of the mid-IR laser through the birefringent crystal by measuring the mid-IR ellipticity after propagation. Due to the crystal birefringnece, the maximum ellipticity obtained in this way was approximately 0.4. For comparison, we also show the measured ellipticity dependence of the 11th harmonic generated for laser polarization in the c-plane of ZnO, for which there is no birefringence. In both cases, the ellipticity dependence behaves similarly. While the harmonic generation efficiency drops substantially at relatively large values of ellipticity, it remains above 50$\%$ for ellipticity values $\leq$0.35, corresponding to crystal orientation angles between $0^{\circ}$-$20^{\circ}$, $70^{\circ}$-$110^{\circ}$, and $160^{\circ}$-$180^{\circ}$. Therefore, we can attribute the discrepancies between theory and experiments close to $45^{\circ}$ and $135^{\circ}$ as resulting from the elliptically-polarized driving laser.


\begin{thebibliography}{xx}
\bibitem {attopulse-1}M. Drescher,  \emph{et al.}, X-ray pulses approaching the attosecond frontier. Science {\bfseries 291}, 1923–1927 (2001).
\bibitem {attopulse-2}P. M. Paul, \emph{et al.}, Observation of a train of attosecond pulses from high harmonic generation. Science {\bfseries 292}, 1689–1692 (2001).
\bibitem {levesque} J. Levesque \emph{et al.}, Phys. Rev. Lett. {\bf 99}, 243001 (2007).
\bibitem {jila} X. Zhou \emph{et al.}, Phys. Rev. Lett. {\bf 102}, 073902 (2009)
\bibitem {Bordeau} Y. Mairesse, \emph{et al.},New J. Phys. {\bf 10}, 025028 (2008).
\bibitem {ATLpol} Anh-Thu Le, R. R. Lucchese, and C. D. Lin, Phys. Rev. A {\bf 85}, 023814 (2010).
\bibitem {ZnO-2011}S. Ghimire, A. D. DiChiara, E. Sistrunk, P. Agostini, L. F. DiMauro, and D. A. Reis, Nat. Phys. {\bfseries 7}, 138 (2011).
\bibitem {chini_apl} S. Gholam-Mirzaei, J. Beetar, and M. Chini, Appl. Phys. Lett. {\bf 110}, 061101 (2017).
\bibitem {Vampa_nature} G. Vampa, T. J. Hammond, N. Thir\'e, B. E. Schmidt, F. L\'egar\'e, C. R. McDonald, T. Brabec, and P. B. Corkum, Nature (London) {\bf 522}, 462 (2015).
\bibitem {Vampa_prl} G. Vampa, T. J. Hammond, N. Thir\'e, B. E. Schmidt, F. L\'egar\'e, C. R. McDonald, T. Brabec, D. D. Klug, and P. B. Corkum, Phys. Rev. Lett. {\bf 115}, 193603 (2015)
\bibitem {ohio_nc} Z. Wang, H. Park, Y. H. Lai, J. Xu, C. I. Blaga, F. Yang, P. Agostini, and L. F. DiMauro, Nat. Commun. {\bf 8}, 1686 (2017).
\bibitem {shambhu_JPB2014}  S. Ghimire,  J. Phys. B: At. Mol. Opt. Phys. {\bfseries 47}, 204030 (2014).
\bibitem {shambhu_OL}  Y. S. You,  \emph{et al.}, Opt. Lett. {\bfseries 42}, 1816 (2017).
\bibitem {T.T.Lu_nature2015} T. T. Luu, M. Garg, S. Yu. Kruchinin, A. Moulet, M. Th. Hassan, and E. Goulielmakis, Nature  {\bf 521}, 498 (2015).
\bibitem {Garg_nature2016} M. Garg, M. Zhan, T. T. Luu, H. Lakhotia, T. Klostermann , A. Guggenmos, and E. Goulielmakis, Nature {\bf 538}, 359 (2016).
\bibitem {shambhu_nc2017} Y. S. You, \emph{et al.}, Nat. Commun.  {\bf 8}, 724 (2017).
\bibitem {Hohenleutne_nature2015} M. Hohenleutner, F. Langer, O. Schubert,M. Knorr, U. Huttner, S.W. Koch, M. Kira, and R. Huber, Nature (London) {\bf 523}, 572 (2015).
\bibitem {graphene_science2017} N. Yoshikawa, T. Tamaya, and K. Tanaka, Science {\bf 356}, 736 (2017).
\bibitem {Taucer_prb} M. Taucer,\emph{et al.}, Phys. Rev. B {\bf 96}, 195420 (2017).
\bibitem {shambhu_mgo} Y. S. You, D. A. Reis, and S. Ghimire, Nat. Phys. {\bfseries 13}, 345 (2017).
\bibitem {schubert_naturephoton2014} O. Schubert \emph{et al.}, Nat. Photonics {\bf 8}, 119 (2014).
\bibitem {Langer_naturephoton2017} F. Langer, M. Hohenleutner, U. Huttner, S. W. Koch, M. Kira, and R. Huber,  Nat. Photonics {\bf 11}, 227 (2017).
\bibitem {shambhu_mos2} H. Liu,  \emph{et al.}, Nat. Phys. {\bfseries 13}, 262 (2016).
\bibitem {berrycurvature-TTLu} T. T. Luu, \&  H. J. W\"orner,  Nature Communications {\bfseries 9}, 916 (2018).
\bibitem {GaSe-PRL}  K. Kaneshima, \emph{et al.}, Physical Review Letters {\bf 120}, 243903 (2018)
\bibitem {Shambhu-MgO-2018}Y. S. You, E. Cunningham, D. A. Reis and S. Ghimire, J. Phys. B, {\bf 51}, 114002 (2018).
\bibitem {Shima} S. Gholam-Mirzaei,  \emph{et al.}, J. Opt. Soc. Am. B {\bf 35}, A27–A31 (2018).
\bibitem {solidargon} G. Ndabashimiye, \emph{et al.}, Nature (London) {\bf  534}, 520 (2016).
\bibitem {A.A.Lanin_optica2017} A. A. Lanin, E. A. Stepanov, A. B. Fedotov, and A. M. Zheltikov, Optica, {\bfseries 4}, 516 (2017).
\bibitem {threestep1} J. L. Krause, K. J. Schafer, and K. C. Kulander, Phys. Rev. Lett. {\bf 68}, 3535 (1992)
\bibitem {threestep2} P. B. Corkum, Phys. Rev. Lett. {\bf 71}, 1995 (1993).
\bibitem {threestep3} M. Lewenstein, \emph{et al.}, Phys. Rev. A {\bf 49}, 2117 (1994).
\bibitem {cdlbook}C. D. Lin,Anh-Thu Le, C.Jin, H.Wei, \emph{Attosecond and Strong-Field Physics: Principles and Applications} (Cambridge University Press, 2018).
\bibitem {ATLQRS} Anh-Thu Le, R. R. Lucchese, S. Tonzani, T. Morishita, and C. D. Lin, Phys. Rev. A {\bf 80}, 013401 (2009).
\bibitem {TDDFT-original}E. Runge and E. K. U. Gross, Phys.Rev. Lett. 52, 997 (1984).
\bibitem {Rubio}Tancogne-Dejean, N., O. D. Mücke, F. X. Kärtner, and A. Rubio ,Phys. Rev. Lett. {\bf 118}, 087403 (2017).
\bibitem {floss} I. Floss, C. Lemell, G. Wachter, V. Smejkal, S. A. Sato, X. M. Tong, K. Yabana, and J. Burgdorfer, Phys. Rev. A {\bf 97}, 011401 (2018).
\bibitem {WMX_pra2015} M. X. Wu, S. Ghimire, D. A. Reis, K. J. Schafer, and M. B. Gaarde, Phys. Rev. A {\bf 91}, 043839 (2015).
\bibitem {Golde_prb2008} D. Golde, T. Meier, and S. W. Koch, Phys. Rev. B {\bf 77}, 075330 (2008).
\bibitem {ttluu-prb}T. T. Luu,  \& H. J. W\"orner,  Phys. Rev. B {\bf 94}, 115164 (2016).
\bibitem {solid-sfa}G. Vampa, C. R. McDonald, G. Orlando, P. B. Corkum, and T. Brabec,  Phys. Rev. B {\bf 91}, 064302 (2015).
\bibitem {TDDFT-open quantum system} Isabella Floss \emph{et al.}, Phys. Rev. B {\bf 99}, 224301 (2019).
\bibitem {jiang_pra} S. C. Jiang, H. Wei, J. G. Chen, C. Yu, R. F. Lu, and C. D. Lin, Phys. Rev. A {\bf 96}, 053850 (2017).
\bibitem {jiang_prl} S. C. Jiang, \emph{et al.},Phys. Rev. Lett. {\bf 120}, 253201 (2018).
\bibitem {comment} We thank Referee B of our previous paper \cite{jiang_prl} for suggesting the LCE model to obtain perpendicular harmonics.
\bibitem {xia_optexpress} Peiyu Xia \emph{et al.}, Opt. Express {\bf 26}, 29393 (2018).
\bibitem {Yao_curvature} Yugui Yao, \emph{et al.}, Phys. Rev. Lett. {\bf 92}, 037204 (2004).
\bibitem {S. W. Koch's book}Haug, H., Koch, S. W. Quantum Theory of the Optical and Electronic Properties of Semiconductors: Fivth Edition. World Scientific Publishing Company, 2009.
\bibitem {Hecht} E. Hecht,  Optics, 4th ed. (Addison-Wesley: Reading, MA, 2001).
\end{thebibliography}
\end{document}